\documentclass[12pt,epsf,amssymb]{article}
\usepackage{epsfig,psfrag}
\usepackage{graphicx}
\usepackage{amsfonts}
\usepackage{amsmath}
\usepackage[usenames]{color}
\usepackage{amssymb}

\def\P{\mathbb{P}}

\begin{document}
\baselineskip 0.6cm
\newcommand{\gsim}{ \mathop{}_{\textstyle \sim}^{\textstyle >} }
\newcommand{\lsim}{ \mathop{}_{\textstyle \sim}^{\textstyle <} }
\newcommand{\vev}[1]{ \left\langle {#1} \right\rangle }
\newcommand{\bra}[1]{ \langle {#1} | }
\newcommand{\ket}[1]{ | {#1} \rangle }
\newcommand{\Dsl}{\mbox{\ooalign{\hfil/\hfil\crcr$D$}}}
\newcommand{\nequiv}{\mbox{\ooalign{\hfil/\hfil\crcr$\equiv$}}}
\newcommand{\nsupset}{\mbox{\ooalign{\hfil/\hfil\crcr$\supset$}}}
\newcommand{\nni}{\mbox{\ooalign{\hfil/\hfil\crcr$\ni$}}}
\newcommand{\EV}{ {\rm eV} }
\newcommand{\KEV}{ {\rm keV} }
\newcommand{\MEV}{ {\rm MeV} }
\newcommand{\GEV}{ {\rm GeV} }
\newcommand{\TEV}{ {\rm TeV} }

\def\diag{\mathop{\rm diag}\nolimits}
\def\tr{\mathop{\rm tr}}

\def\Spin{\mathop{\rm Spin}}
\def\SO{\mathop{\rm SO}}
\def\O{\mathop{\rm O}}
\def\SU{\mathop{\rm SU}}
\def\U{\mathop{\rm U}}
\def\Sp{\mathop{\rm Sp}}
\def\SL{\mathop{\rm SL}}

\def\change#1#2{{\color{blue}#1}{\color{red} #2}\color{black}\hbox{}}


\begin{titlepage}

\begin{flushright}
LTH/732
\end{flushright}

\vskip 2cm
\begin{center}
{\large \bf Metastable Vacua, Geometrical Engineering}\\                                                   
{\large \bf and MQCD Transitions}
\vskip 1.2cm
Radu Tatar and Ben Wetenhall

\vskip 0.4cm

{\it Division of Theoretical Physics, Department of Mathematical Sciences

The University of Liverpool,
Liverpool,~L69 3BX, England, U.K.

rtatar@liverpool.ac.uk,~~benweten@liv.ac.uk}

\vskip 1.5cm

\abstract{ We consider the geometrical engineering of the non-supersymmetric metastable 
vacua of ${\cal N} = 1$ super Yang-Mills proposed in \cite{iss} and \cite{agava1}. By T-duality they become 
${\cal N} = 1$ brane configurations. The identifications between gluino condensation and the geometry sizes 
for the configurations  of \cite{agava1}
are studied by proceeding through the usual MQCD transitions. The geometrical description of the Seiberg dualities 
for the theories of \cite{iss} involves new types of modifications of the complex structure for the resolutions of
${\cal N} = 2$ singularities.
} 

\end{center}
\end{titlepage}


\section{Introduction}

The study of wrapped D-branes on cycles of Calabi-Yau manifolds has provided many insights into the study of 
gauge theory dynamics. The geometrical set-up with wrapped D5-branes and the brane configurations with 
D4-branes and NS branes are T-dual to each other. The two approaches provide the same type of information about the underlying field theories.
There are advantages and disadvantages for either of the two approaches:

1) the IIA brane configuration picture can be easily lifted to an M-theory configuration which describes the strongly coupled
regime of the field theory (for a review see \cite{giku}). The disadvantage is the lack of a SUGRA solution.

2) there is a SUGRA solution in IIB for the wrapped D5-branes on 2-cycles of the resolved conifold, in the presence of an NS flux 
\cite{bdt1}-\cite{bdt6}. The disadvantage is that the field theory contains some extra flavors (D7-branes which are moved to 
infinity).

Recently it was realized that 4-dimensional ${\cal N}=1$ Super Yang-Mills theories with massive matter admit
non-supersymmetric meta-stable vacua \cite{iss}. Subsequently, much work has been dedicated to obtaining a string theory 
picture for these new vacua \cite{fu}-\cite{argurio}~(other field theory directions were explored in \cite{Forste}-\cite{Fischler}). 

Metastable non-SUSY vacua can also appear in string theory by considering systems of D-branes and anti D-branes. Systems of 
D5-branes - anti D5-branes were considered in the work of \cite{agava1} and  configurations with D5-branes and anti D3-branes 
were used in \cite{verlinde}. Other metastable brane configurations with wrapped branes were considered in \cite{diaconescu}.
The configuration of \cite{agava1} contains D5 branes and anti D5-branes wrapped on different 2-cycles. Because the 
geometry is rigid, they cannot cancel each other. The geometrical transition \cite{vafa,civ} 
holds in the presence of the anti D5-branes as one can identify the gluino condensates in field theory 
with the sizes of the $S^3$ cycles in the deformed geometry. 

Our goal is to study the metastable vacua discussed in \cite{iss} and \cite{agava1}. 
In section 2, the geometrically engineered configurations of the type
\cite{agava1} are translated into a brane configuration picture and the corresponding MQCD transition is similar to the 
ones of \cite{dot1}-\cite{dot4}. The (anti)D5 branes are mapped into (anti)D4-branes on separated intervals
between NS branes, which prevents their annihilation. There are two types of geometries that one consider. The first type is when all 
the wrapped ${\bf P^1}$ cycles are in the same homology class and this was the case studied in \cite{agava1} \footnote{We would like 
to thank Cumrun Vafa for helping us clarify this issue.}. The second type is when the wrapped  ${\bf P^1}$ cycles are in different 
homology classes. The stable configuration of D4-branes and anti D4-branes for the latter was discussed in \cite{Mukhi}.

After lifting to M-theory, the corresponding M5 brane will have several 
disjoint parts and each levels out into planar M5 branes which reduce to disjoint deformed conifold singularities. The sizes of the corresponding
3 cycles $S^3$ are then identified with the gluino condensates on the D5-branes and anti D5-branes respectively.  This constitutes 
an important test of the new type of geometric transition introduction in \cite{agava1}.

In section 3 we consider the geometrical picture dual to the brane configurations used in \cite{oo2,franco,bena}. The geometry with massive flavors and
the corresponding Seiberg duality have been discussed extensively before in \cite{dot2,dot5}. The wrapped D5-branes on cycles of resolved conifold geometries
were used to consider Seiberg dualities as flops in the geometry \cite{plesser,hanany,cachazo}. The procedure of \cite{dot2}  
can be applied for the case when the flavors are either massless or are integrated out 
\footnote{The Seiberg duality for very massive flavors in the IIA brane configuration was considered in \cite{sundrum}.}.   

When the masses of the flavors are smaller than the scale the situation becomes trickier. 
Seiberg duality is a quantum symmetry and its full description should be clear in 
either M-theory or F-theory. In brane configurations we only expect a classical equivalence. In the work of \cite{giku1} the Seiberg duality 
 was considered as a classical 
equivalence between Higgs branches and their deformations. The brane configurations were used to relate the Higgs branches of the 
gauge groups $U(N_c)$ and $U(N_f-N_c)$. After going to the Higgs phase, there is a freedom of moving the branes by turning on a D-term for the 
$U(1)$ of either  $U(N_c)$ or $U(N_f-N_c)$. The moduli space of the electric and magnetic theories provide different descriptions for the same moduli
space of brane configurations. 

One could consider the brane configuration of the Seiberg duality for massless flavors and then deform the electric and magnetic theories by giving 
masses and expectation values, respectively. But it gets harder to visualise this change in the brane configurations for more complex theories.
A unified description of getting the metastable vacua is required to deal with all possible theories. 
The present work is a first step towards reaching this goal.
The method we propose gives a clearer picture of the different vacua of the magnetic theory and also 
points to the origin of the different branches in the magnetic theory vacua. 

In this work we develop the following procedure (valid for the limit of very light massive quarks 
considered in the metastable vacua approach):

- Consider the resolved conifold with the color D5 branes wrapped on the compact ${\bf P^1}$ cycle and the flavor D5 branes wrapped on some 
non compact  holomorphic 2-cycle.  

- Take a very small  non-holomorphic deformation of the  ${\bf P^1}$ cycle such that it touches the non compact  holomorphic 2-cycle. 
By simultaneously rotating one of the line bundles such that  it touches the North Pole of the  ${\bf P^1}$ cycle, 
the result of these deformations  is a holomorphic cycle in a new complex structure. The rotation of the line bundle determines a rotation of 
the non compact  holomorphic 2-cycle which aligns with the compact cycle.  

- In the new complex structure we are now ready to go through the procedure of  \cite{giku1} and perform the Seiberg duality as a flop in the new geometry. 
The flop does not change the complex structure. In the Seiberg dual geometry, the  flavor branes and color branes are still aligned.

- If we want to have the magnetic theory in the original complex structure, we need to deform back the  ${\bf P^1}$  and  
rotate the line bundle back to  the original position. In this case, the magnetic flavor non compact 2-cycle remains unchanged when the line bundle is 
rotated as it does not end on the rotating line bundle anymore. It was a holomorphic cycle in the deformed complex structure but it is clearly 
non holomorphic in the original complex structure.

When the cycles are rotated, there is a 
tachyon mode between the various wrapped branes. One way to circumvent this tachyonic mode is to allow the bounding of D5-branes which implies 
a recombination of the geometrical cycles. The  ${\bf P^1}$ cycle changes into a non compact holomorphic cycle ending on the rotated line
bundle and is accompanied by a  non compact non holomorphic cycle ending on the non rotated line bundle.
 
The above manipulations are a sign that more general deformations for $A_n$ singularities are required in order to handle the geometries of 
metastable vacua. We comment on this at the end of Section 3 and leave the details for  future work.

\section{Metastable vacua with Branes and anti-Branes}
\label{sec:BottomUp}
In what follows we will use the following directions for the branes:

- in type IIA, the brane configurations contain an NS brane in the directions (012345), an NS' brane in the (012389) directions and 
D4-branes in the (01237) directions.

- in type IIB, the wrapped branes are D5-branes in the direction (012367) where $x^6$ is the angular direction of the $S^2$.

- in the M-theory discussion we use the following notations:~~$v~=~x_4~+~i~x_5,~w~=~x_8~+~~i~x_9$ and 
$t=\mbox{exp}(-R^{-1}x_7~-~i~x_{10})$ where $R$ is the radius of the circle $S^1$ in the 11-th direction.

In the recent paper \cite{agava1} it has been discussed that not only the wrapped D5-branes can be studied during the geometric transitions, 
but also anti D5-branes. The usual geometric transition can be seen as replacing 
wrapped D5-branes on two cycles $P^1$ by fluxes on 3-cycles $S^3$. The wrapped D5-branes correspond to the UV limit of the
field theory and the fluxes to the IR limit of the field theory. The mapping requires the identification of the 
number $N_k$ of wrapped D5-branes (the rank of the gauge group) with the flux of the $H_{RR}$ 3-form through the 
$S^3_k$ as 
\begin{equation}
\label{rankgroup}
\int_{S^3_k} H_{RR} = N_{k}
\end{equation}
and the identification of the gluino condensate in the field theory with the size  of the 3-cycle $S^3_k$
\begin{equation}
\int_{S^3_k} \Omega^{(3,0)} = S_k.
\end{equation}
The new ingredient of \cite{agava1} was to consider some extra anti D5-branes wrapped on 2-cycles. This extends 
the equation (\ref{rankgroup}) to negative numbers and the conjecture of \cite{agava1} is that the geometric
transition duality also holds for systems of D5-branes and anti D5-branes. 

We can reformulate the new conjecture in terms of type IIA brane configuration by using the 
results of \cite{dot1}-\cite{dot4}. The wrapped D5-branes on the $S^2$ can be mapped 
into D4-branes on the interval given by the radial direction of the 
$S^2$. The singular lines inside the resolved conifold are mapped into a pair of orthogonal 
NS branes. 

What happens if one wraps anti D5-branes? They are mapped into anti-D4 branes lying between two 
orthogonal NS branes.  If we have both wrapped D5-branes and wrapped anti D5-branes, the system will be 
mapped into D4-branes and anti-D4 branes. This  type of configuration has been extensively 
discussed in  the work of \cite{Mukhi} \footnote{A similar system has been considered in \cite{radu} , but with D5-branes and anti D5-branes 
which are wrapped  on the vanishing cycle of a singular conifold.}.  

Let us consider a resolved geometry with many $S^2$ cycles and wrap $N_k$ D5-branes on the $k$-th cycle and
$N_{k'}$ anti D5-branes on the $k'$-th cycle. We now distinguish between two cases:

1) The ${\bf P^1}$ cycles are in the same homology class which is the case considered in \cite{agava1}. 
The geometry is obtained by starting with a resolved  ${\cal N} = 2, A_1$ singularity and then deforming by adding 
\begin{equation} 
W = \sum_{k=1}^{n+1} \frac{g_k}{k} \mbox{Tr} \Phi^k
\end{equation}
where $\Phi$ is the unrestricted direction in the normal bundle. We get a collection of $n$ resolved conifolds  ${\cal N} = 1$ singularity which 
contains $n$ $P^1$ cycles in the same homology class. 

We can also add D5 branes on each of the ${\bf P^1}$ cycles. 
After a T-duality this will become a straight NS brane and a curved NS' brane, as discussed in \cite{dot2}, see Figure 1. 

\begin{figure}
\centerline{\epsfxsize=60mm\epsfbox{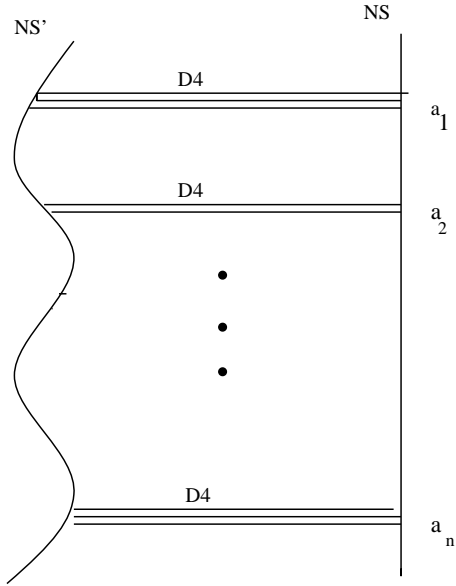}} \caption{The T-dual configuration of D5 branes distributed among $n$ $\P^1$.} 
\end{figure}

In the limit $g_N \rightarrow \infty$ this changes into Figure 2, where the curved NS' changes into $n$ straight NS' branes orthogonal on the NS brane.

\begin{figure}
\centerline{\epsfxsize=60mm\epsfbox{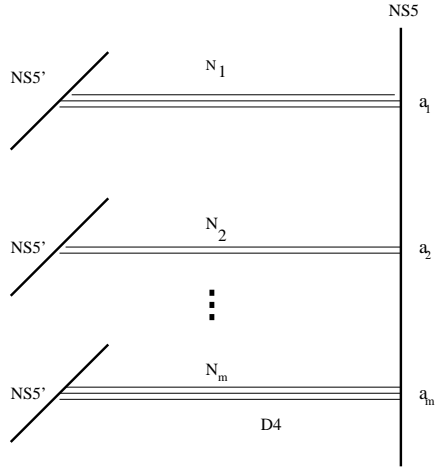}} \vspace*{1cm} \caption{The brane configuration in the limit
$g_n \to \infty$.} 
\end{figure}

To reach the configuration of \cite{agava1} we replace some stacks of D5 branes with stacks of anti D5 branes. By starting with $N_k$ D5-branes 
on the $k$-th cycle and $N_{k'}$ anti D5-branes on the $k'$-th cycle, after T-duality we get $N_k$ D4-branes at $a_k$ and 
$N_{k'}$ anti D4-branes at  $a_{k'}$. 

2)  The ${\bf P^1}$ cycles are not in the same homology class. In this case one starts with the resolution of  ${\cal N} = 2, A_n$ singularity and wrap 
D5 branes on each of the  ${\bf P^1}$ cycles. The T-dual is a configuration with D4-branes between pairs of parallel NS branes:

\begin{figure}
\centerline{\epsfxsize=100mm\epsfbox{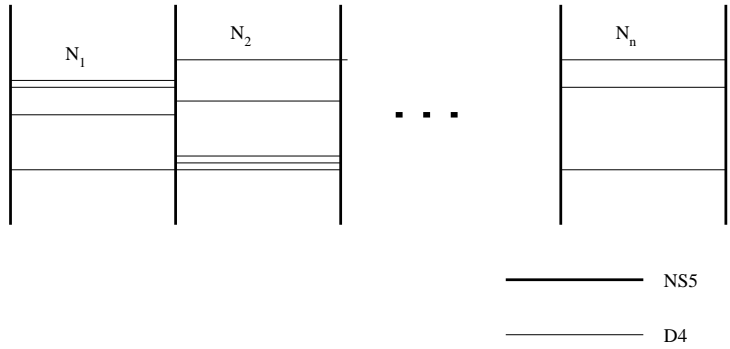}} \vspace*{1cm} \caption{$A_n$ brane configuration: D5-branes 
wrapping $\P^1$ cycles are T-dualized to D4 branes between NS branes.} 
\end{figure}

By adding masses for the adjoint fields, one gets an ${\cal N} = 1$ configuration with D4-branes between rotated NS branes. One can replace 
some of the D4-branes with anti D4-branes. In order to reduce the discussion to the one of the previous case, 
let us consider that we have $N_k$ D4-branes between the $k$-th NS brane and the $k+1$-th NS brane and $N_{k'}$ anti D4-branes 
between the $k'$ NS brane and the $k'+1$ NS brane. 

The result of \cite{Mukhi} for $k+1=k'$ is that the two D4-branes and anti D4-branes
repel each other when adjacent. Nevertheless,~if the D4-branes and anti D4-branes are not adjacent, 
the brane configuration then becomes stable and 
the stability also holds in the T-dual picture with wrapped D5-branes and anti D5-branes.

The IIA configuration can then be lifted to M-theory and one can then go through the MQCD transition of \cite{dot1}.
As the configuration of D4-branes and NS branes is lifted to a single M5 brane, the same holds for the 
configuration of anti D4-branes and NS branes. We start with case 2) which is simpler to describe and the we return to case 1). 

Remember that the D4-NS system is lifted to an M5 brane:
\begin{equation}
\label{m5d4}
v^{N_{k}}=t,~~w^{N_k}=\xi^{N_k} t^{-1},~~v w = \xi
\end{equation} 
where $\xi$ is related to $\Lambda$, the dynamical scale.

The $N_k$ D4-branes have the orientation as starting from the NS brane extended in the $v$-direction
(therefore we have $v^{N_{k}} = t$) and ending on the NS brane extended in the $w$ direction 
(therefore we have $w^{N_k}=\xi^{N_k} t^{-1}$). For the case of $N_{k'}$ anti D4-branes between 
an NS brane ($v$ direction) and an NS' brane ($w$ direction) the situation changes as the anti D4-brane pulls (starts on) 
the NS' brane and pushes (ends on) the NS brane. The corresponding M5-brane is 
\begin{equation}
\label{m5ad4}
w^{N_{k'}}=t,~~v^{N_{k'}}=\xi_1^{N_{k'}} t^{-1},~~v w = \xi_1.
\end{equation} 
       
If one then closes the $S^2$-cycles that the D5-branes are wrapped on, this corresponds to closing the 
intervals between the  $k$-th NS brane and the $k+1$-th NS brane and between  the $k'$-th  NS brane 
and $k'+1$-th NS brane, respectively. The result is that the M5 brane (\ref{m5d4}) becomes a 
collection of $N_{k}$ planar M5 branes \cite{dot1}:
\begin{equation}
\label{m5d4p}
\Sigma_l:~~t=t_0,~~v~w = \xi~\mbox{exp} (2\pi i l/N_{k}),~l=0,\cdots,N_{k}-1
\end{equation}

The M5 brane (\ref{m5ad4}) has a similar form, the only difference being that $\xi$ and $N_k$ are replaced by $\xi_1$ and $|N_{k'}|$:
\begin{equation}
\label{m5ad4p}
\Sigma_n:~~t=t_1,~~v~w = \xi_1~\mbox{exp} (2\pi i l/|N_{k'}|),~l=0,\cdots,|N_{k'}|-1
\end{equation}
where $|N_{k'}|$ is the absolute value of the flux due to the $k'$ anti D4-branes.

The only question is how are $\xi$ and $\xi_1$ related. For the D4-branes the value of $\xi$ is  
\begin{equation}
\xi = \Lambda_0^3 \mbox{exp}(-\frac{2 \pi~i~\alpha}{N_k}) 
\end{equation} 
where $\Lambda_0$ is the cut-off scale and $\alpha$ is the bare coupling constant:
\begin{equation}
\label{d4}
\alpha = - \frac{\theta}{2 \pi} - i \frac{4 \pi}{g_{YM}^2}.
\end{equation}
The Yang-Mills coupling constant is written in terms of the geometry and string constants as
\begin{equation}
\label{cc}
\frac{1}{g_{YM}^2}=\frac{\Delta L}{g_s l_s}
\end{equation} 
where the value of $\Delta L$ measures the distance between the two NS branes. We can ask what is the difference between having
D4-branes or anti D4-branes. The measurement of $\Delta L$ is in opposite direction for anti D4-branes as 
compared to the D4-branes. Therefore, if we measure   $\Delta L$ from left to right, it results that if for a D4-brane $\alpha$ is given
by (\ref{d4}), the corresponding  coupling constant for anti D4-branes is $\bar{\alpha}$. 

This implies that for anti D4-branes the value of $\xi_1$ is 
\begin{equation}
\xi_1 = \Lambda_0^3~\mbox{exp}~(-\frac{2 \pi~i~\bar{\alpha}}{|N_{k'}|}). 
\end{equation}

The main result of \cite{dot1} was that, after the MQCD transition, the 
value of $\xi$ was related to the size of the $S^3$ in the deformed geometry. 
But equation (\ref{m5d4p}) reduces exactly to the deformed conifold when reducing from M theory to type IIA, with the size of the 
deformation $S^3$ being  $\xi \mbox{exp} (2\pi i l/N_{k})$. 
Because of the above relation between 
$\xi$ and $S_k$, we see that the geometric transition conjecture holds if the gluino condensate for the 
gauge group on the D4 branes (identified with the size of the $S^3_{k}$) is 
\begin{equation}
<S_k> = \Lambda_0^3~\mbox{exp}(-\frac{2 \pi i \alpha}{N_k})~\mbox{exp} (2\pi i l/N_{k}),~l=0,\cdots,N_{k}-1 .
\end{equation} 

The same thing holds for anti D4-branes. The curve (\ref{m5ad4p}) reduces to a deformed conifold with the 
size of the $S^3_{k'}$) being $\xi_1 \mbox{exp} (2\pi i l/|N_{k'}|)$. There is similarity between the 
deformation of the geometry with cycles with positive and negative fluxes. The relation between   $\xi_1$ and 
$S_{k'}$ implies that 
\begin{equation}
<S_{k'}> = \Lambda_0^3~\mbox{exp}(-\frac{2 \pi i \bar{\alpha}}{N_{k'}})~\mbox{exp} (2\pi i l/|N_{k'}|),~l=0,\cdots,N_{k'}-1.
\end{equation}

We can now go to the case 1), in the $g_N \rightarrow \infty$ limit. The are D4-branes and anti D4-branes ending on the NS brane
at $a_k$ and $a_{k'}$. For the case of two stacks of D4 branes, the M5 brane would have the form:
\begin{equation}
t = (v-a_k)^{N_k} (v-a_{k'})^{N_{k'}}, w = \frac{\xi_1}{v-a_k} +   \frac{\xi_2}{v-a_{k'}}
\end{equation}
where $\xi_i$ are equal to $\Lambda_i^3$ with $\Lambda_i$ being the dynamical scales of the ${\cal N} = 1$ theories. The  $\Lambda_i$ are related to 
the ${\cal N} = 2$ scales by threshold condition
\begin{equation}
\Lambda_k^{3} = g_{n+1}
\Lambda_{{\cal N}=2}^{2 N/N_k} (a_k - a_{k'})^{1 - 2 N_{k'}/N_k};~\Lambda_{k'}^{3} = g_{n+1}
\Lambda_{{\cal N}=2}^{2 N/N_{k'}} (a_{k'} - a_{k})^{1 - 2 N_{k}/N_{k'}}
\end{equation}
This formula is obtained after integrating out the massive adjoint field of mass $g_{n+1} (a_k - a_{k'})$ and the massive W-bosons 
of mass $(a_{k'} - a_{k})^{- 2 N_{k}/N_{k'}}$. The mass of the massive adjoint field is unchanged by 
replacing the D5-branes with anti D5-branes but the W boson masses change. The change is due to the change of orientation of 
the D5 branes into anti D5 branes. There is also a change $\alpha_2 \rightarrow \bar{\alpha_2}$ as discussed before.
This two changes explain the replacement of the equation (3.13) of \cite{agava1} with their equation (3.14). 

In \cite{agava1} it is shown an explicit identification between the field theory  and geometrical quantities. As discussed in \cite{dot2},
the problems which arise in making the same identifications in the MQCD transitions are due to the fact that the geometrical curve is hyperelliptic and the 
MQCD curve is rational. The only case when the identification can be made is in the case of quadratic superpotentials for the 
adjoint field, which reduces to the case 2).
 
After discussing the metastable vacua from systems of branes and antibranes, we consider the case of metastable vacua from 
rotated branes in the next section. 

\section{Metastable Vacua with Branes at Angles}
\label{sec:Angles}

In this section we consider the 
metastable vacua discussed in \cite{iss}. The brane configuration and the MQCD picture 
have been considered in  \cite{oo2,franco,bena}. The work of \cite{bena} arrived at a negative conclusion in  
concerning the possibility of having an MQCD picture for such metastable vacua. We will argue that 
a more general framework of deformations of $A_n$ singularities might be needed in order to 
obtain such an MQCD picture.  

We start with a very brief review of the field theory results. We have an ${\cal N} =1~~SU(N_c)$ theory with 
$N_f$ massive flavors $Q, \tilde{Q}$, their mass being much smaller than the dynamical scale $\Lambda$. We work in the 
range $N_c + 1 \le N_f < \frac{3}{2} N_c$, such that the magnetic  phase is free. 

The Seiberg dual is ${\cal N} =1~~SU(N_f-N_c)$ with $N_f$ flavors $q, \tilde{q}$ and the meson $M$ together with a 
superpotential
\begin{equation}
\label{dualpot}
W = h Tr(q M \tilde{q}) - h \mu^2 Tr(M).
\end{equation}
The region where the masses are very small is characterized by a breaking of SUSY due to the F-term of $M$ which implies that  
$q, \tilde{q}$ have $N_f - N_c$ nonzero vacuum expectation values which equal the $N_f - N_c$ largest masses of 
the electric theory. 

In terms of brane configurations, the electric picture contains the same NS, NS' and 
electric D4-branes as in the previous section but also some semi-infinite D4 branes ending on the NS or NS' branes. 
For the D4-branes ending on the NS brane, the distance between the $N_c$ gauge D4-branes
and the semi-infinite flavor D4-branes is the mass of the flavors \footnote{This is due to the fact that the 
${\cal N}=1$ brane configuration comes from an ${\cal N} = 2$ configuration by rotating the NS' brane. The direction 
of the NS brane describes the Coulomb branch of the ${\cal N} = 2$ theory (the vacuum expectation value of the scalar field) so 
by moving on the Coulomb branch one gives a vacuum expectation value to the adjoint field $\Phi$ and a mass to the fundamental quarks due to the 
coupling $Q \Phi \tilde{Q}$.}. For the D4-branes ending on the NS' brane, the distance between the $N_c$ gauge D4-branes
and the semi-infinite D4-branes is the vacuum expectation value of the meson $M$ .

The brane configuration is shown in Figure 4. The angle $\theta$ is related to the mass of the  ${\cal N} = 2$ adjoint field by
$\mbox{tan} \theta = \mbox{mass}$. In what follows we will take $\theta = \pi/2$, i.e. the case when the adjoint field is infinitely massive.
The case with $\theta \ne \pi/2$ is also interesting as it would describe the metastable vacua considered in  \cite{A2}.

\begin{figure}[ht]
\begin{center}
\psfrag{a}{D$4_m$}
\psfrag{b}{D$4_M$}
\psfrag{c}{D$4_c$}
\psfrag{m}{$m$}
\psfrag{M}{$M$}
\psfrag{al}{$\theta$}
\psfrag{e}{$89$}
\psfrag{f}{$7$}
\psfrag{g}{$45$}
\psfrag{k}{${\it NS5}$}
\psfrag{l}{${\it NS5}'$}
\epsfig{file=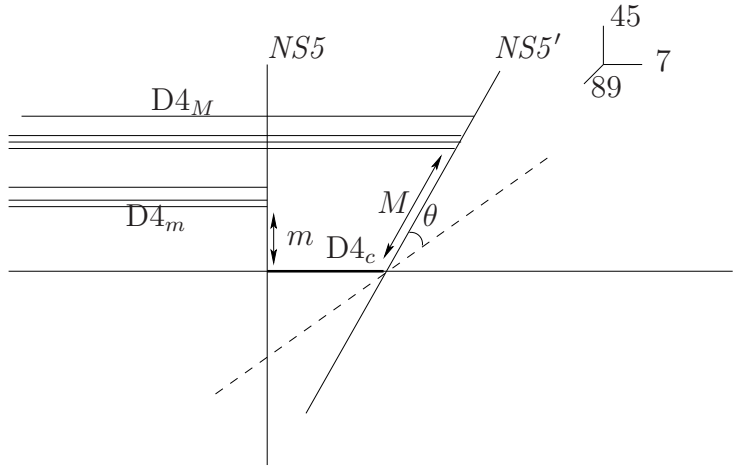,height=6cm}
\caption{ Brane construction .\label{branes}}
\end{center}
\end{figure}

After a Seiberg duality, the position of the NS and NS' branes interchange. In considering the moduli space of vacua, there is a big difference between 
having massive quarks with a mass bigger or a mass lower than the scale, in both the electric and magnetic pictures. 
If the masses are bigger than the scale, the quarks are integrated out 
and what we get is an electric scale
\begin{equation}
\tilde{\Lambda}_e^3 = \Lambda_e^{{3 N_c - N_f}/N_c}(\prod \mu)^{1/N_c}
\end{equation}
and a magnetic scale
\begin{equation}
\tilde{\Lambda}_m^3 = \Lambda_m^{{3 \tilde{N}_c - N_f}/\tilde{N}_c}(\prod \mu)^{1/\tilde{N}_c}
\end{equation}
where $\tilde{N}_c=N_f-N_c$. 
As considered in \cite{dot2}, the lift of this brane configuration to M-theory is an M5 brane depending on 
$\mu$. 

For the non-SUSY vacuum described above, the position of the NS and NS' branes interchange but the exchange between the masses of the electric quarks 
and those of the magnetic quarks does not hold. This is because the magnetic quarks are massless but have vacuum expectation values instead.
The mass of the electric quarks is mapped into the vacuum expectation value of the magnetic quarks. Because the mass of the magnetic quarks is measured by distances on 
the NS' brane, the vevs of the magnetic quarks are measured on the NS brane. 

We now want to explicitly perform a Seiberg duality. In the brane configuration language, for the case of massless electric 
quarks this has been shown explicitly in 
\cite{giku1}. As the stack of the $N_f$ electric flavor branes touches the stack of the $N_c$ electric color branes, we can bind 
together $N_c$ color branes with $N_c$ flavor branes and move them together with the NS' brane in the $x^6$ direction, then in 
the $x^7$ direction and back in the $- x^6$ direction. What we get is the dual theory, with $N_f - N_c$ color D4 branes and 
$N_f$ flavor D4-branes.

There is no such explicit construction for the case of light flavors because there is no way to bind color D4 branes to 
light flavor D4-branes without breaking SUSY in the electric theory.

The displacement on the NS brane direction is the same in both electric and magnetic pictures. In the magnetic picture,
the flavor D4-branes are at an angle with respect to the color D4-branes.
The angle is given by
\begin{equation}
\label{angle}
\mbox{tan}(\alpha)=\frac{\mu}{\Delta L}
\end{equation}
where $\Delta L$ is the length of the color D4 branes and is related to the field theory coupling constant $g$ by
\begin{equation}
\frac{1}{g^2}=\frac{\Delta L}{g_s l_s}.
\end{equation}

\subsection{First Choice: Geometry and NS flux}

One  way to obtain branes at angles in type IIA brane configurations is to start with IIB branes and NS flux.
Let us consider the following starting point in type IIB: take finite D5-branes wrapped on an $S^2$ with coordinates $(y,~\theta_2)$ and 
infinite D5-branes wrapped on a non-compact cycle with same radial and angular coordinates. 
Add some $B_{NS}$ field in $(y,\theta_1)$ directions on both the finite $S^2$ and the non-compact 2-cycle, where $\theta_1$ is a direction
orthogonal to the $S^2$ cycle
\footnote{This is different from the discussion of \cite{radu} where there were pairs of D5-branes and anti D5-branes on the vanishing cycle of 
a conifold. The $B_{NS}$ was turned on both directions of the $S^2$ cycle and the  D5 brane and anti D5-brane pair gave rise to an integer 
D3-brane.}.

A T-duality in the $y$ direction takes the D5-branes into D4-branes which are still parallel.  
It is well-known that a T-duality on the $y$  direction of the $T^2(y,\theta_1)$ in the presence of a $B_{NS}$ field
determines a rotation of the $T^2$ by an angle
\begin{equation}
\mbox{tan} \beta =  B_{NS}(y,\theta_1)
\end{equation}
This means that the coordinates  $(y,\theta_1)$ are rotated into  $(y',\theta'_1)$ by an angle $\beta$. 

Denote the coordinates of the NS branes and the D4-branes by $(r,z,y',\theta'_1,x,\theta_2)$,~the NS branes being extended
in the directions NS($x,\theta_1)$ and NS'($z,r$). We rotate the direction 
$\theta'$ till it coincides back with $\theta_1$. 
The semi-infinite D4 brane do not feel the effects of the rotations, as they are extended in the $\theta_2$ direction.
Hence the introduction of the NS field does not give the wanted picture with rotated D4 branes.

\subsection{Second Choice: Just Geometry}

Consider the resolved conifold. The small resolution is covered by two copies of $C^3$ with coordinates $Z,X,Y (Z',X',Y')$.
The resolved conifold geometry is
\begin{equation}
\label{embedding}
Z'=1/Z,~~X' = X Z,~~Y'=Y Z
\end{equation}
which has a compact 2-cycle $Z'=1/Z$. If we wrap D5-branes on the 
compact 2-cycles we get a gauge group on the D5-branes.

We can also define non compact holomorphic cycles. To do so, we actually start with an ${\cal N} = 1$ deformed $A_3$ singularity which, after resolution, 
gives a collection of 3 resolved conifold geometries 
\begin{equation}
Z_i'=1/Z_i,~~X_i' = X_i Z_i,~~Y_i'=Y_i Z_i,~~i=1,2,3
\end{equation}
where $X_1=X_2',~X_2=X_3'$, $Y_1=Y_2',~Y_2=Y_3'$.
 
We have three compact 2-cycles given by $Z_i'=1/Z_i,~i=1,2,3$. We can keep the second  2-cycle compact i.e. we keep the 
lines $X_1=X_2'$ and $X_2=X_3'$ unchanged. At the same time we take the lines $X_1', Y_1'$ and $X_3, Y_3$ to infinity which means that the
compact left and right 2-cycles become holomorphic non-compact cycles. In what follows we denote 
\begin{equation}
X_1=X_2'=X', Y_1=Y_2'=Y'~~~\mbox{and}~~X_2=X_3'=X,~Y_1=Y_2'=Y.
\end{equation}

The non compact 2-cycles we are going to considered in this work are
\begin{equation}
\label{ncom}
Y=0,~~X = \mbox{mass}~~~\mbox{or}~~~Y'=0,~~X' = \mbox{vev}.
\end{equation}

As discussed in \cite{dot2}, the Seiberg duality can be obtained by a birational flop in the geometric engineering. For the 
resolution of the conifold, this means an exchange of the role of $X(Y)$ and $Y'(X')$ in the resolution, together with
interchanging the left and right non compact 2-cycles. This flop appears quite natural for massless flavors.
For the massive flavors, the flop still exchanges the $X(Y)$ and $Y'(X')$ in the resolution but 
it is less clear on how to handle the flavors.  
The metastable solution of \cite{iss} corresponds to light flavors in the electric theory and massless flavors in the magnetic theory but with 
vacuum expectation values. 

We still want to view the Seiberg duality as a flop in some geometry, where the color D5-branes and the flavor D5-branes touch each other.
To do this, we need to perform changes in  the geometry such that it resembles the one considered in \cite{giku1}. We consider 
a non holomorphic deformation of the  ${\bf P^1}$ cycle, together with a rotation of the line bundle,  in the following way:

$\bullet$ move the North Pole of the  ${\bf P^1}$ cycle along the direction $X'$ by a very small distance $\mu$. 
The projection from the North Pole, used to define 
the coordinate $Z'$, changes and this makes the transition function from the upper to lower coverings non-holomorphic.
In terms of brane configurations, this means a rotation of the D4 branes by an angle (\ref{angle}).

$\bullet$ the axis $X'$ is then rotated  by the same angle (\ref{angle}) until it  becomes tangent to the North pole of the  ${\bf P^1}$ cycle.
The value $\mu$ in (\ref{angle}) is very small so the angle $\alpha$ is also very small. 

$\bullet$ the non compact 2-cycle ending on $X'$ is also forced to rotate by  (\ref{angle}) and in the final configuration there is an 
alignment between the two stacks of D5 branes.  There is a map between the initial and final holomorphic transition functions 
\begin{equation}
\label{map}
Z' = 1/Z~~~\rightarrow~~~\tilde{Z}' = 1/\tilde{Z}.
\end{equation}
In the geometry  $C^3(\tilde{X},\tilde{Y}=Y,\tilde{Z})$, the North pole of the ${\bf P^1}$ coincides with the 
South Pole of the infinite holomorphic 2-cycle. 

The normal bundle also changes in order to describe the proper
embedding of the ``new'' ${\bf P^1}$ into the resolved conifold. The overall change is 
\begin{equation}
\label{change}
C^3(X',Y',Z')~~~\rightarrow~~~C^3(\tilde{X},\tilde{Y}=Y,\tilde{Z}).
\end{equation}
The line bundle $X'$ rotates into the line bundle $\tilde{X}$, whereas the line bundle $\tilde{Y}$ remains unchanged. 
The size of the  ${\bf P^1}$ cycle remains the same. This means that the gauge coupling constant does not change after the 
geometrical manipulations, as it should. 

As shown in \cite{dot1}, there is a 1-1 map between the geometrical coordinates and the MQCD coordinates when the brane configuration
is lifted to 11 dimensions. This is 
\begin{equation}
Z \leftrightarrow t,~~X' \leftrightarrow v,~~Y \leftrightarrow w
\end{equation}
We can try to see the change (\ref{change}) in the MQCD coordinates. Start with the usual M5 brane wrapped on a holomorphic curve, in the presence
of massive matter:
\begin{equation}
\label{m5}
t~=~w^{N_c-N_f}(w~-~(\frac{\Lambda_{{\cal N}=1}^{3N_c-N_f}}{\mu^{N_c-N_f}})^{1/N_c})^{N_f}~~,~~v w~=~(\Lambda_{{\cal N}=1}^{3N_c-N_f}\mu^{N_f})^{1/N_c} 
\end{equation}
which can be rewritten in terms of only $t$ and $v$ as
\begin{equation}
v^{N_c}~t = \Lambda_{{\cal N}=1}^{3N_c-N_f} (\mu-v)^{N_f} .
\end{equation} 
Now we perform the change in the complex structure by very small rotations of angle (\ref{angle}) in the $(x^4,x^7)$ plane. 
When rotating the  ${\bf P^1}$ cycle, the radial direction will not change in the limit of small angles  (i.e. small quark masses).
If the origin of the $\tilde{X}', \tilde{Z}$ is chosen at the new point of intersection, a rotation of the line bundle takes us to
\begin{equation}
\tilde{v}^{N_c} \tilde{t} =  \Lambda_{{\cal N}=1}^{3N_c-N_f} \tilde{v}^{N_f}
\end{equation} 
in the limit of very small $\mu$ and $\tilde{v} \rightarrow \infty$. But this is just the usual asymptotic NS region
\begin{equation}
\tilde{v} \rightarrow \infty, w \rightarrow 0,~~\tilde{t} \rightarrow \Lambda^{3 N_c - N_f} \tilde{v}^{N_f - N_c} .
\end{equation}
In the case of very small masses, the asymptotic regions $v \rightarrow \infty$ and $ \tilde{v} \rightarrow \infty$ are identical so the small deformation of the 
complex structure is invisible.    

The coordinate $w$ is unchanged by the above manipulations. The relation between the coordinate $\tilde{t}$ and $w$ is similar to the one in 
(\ref{m5}) and this tells us that the usual asymptotic NS' region is obtained
\begin{equation}
w \rightarrow \infty, \tilde{v} \rightarrow 0,~~\tilde{t} \rightarrow w^{N_c} .
\end{equation}

The curve obtained is then similar to one with massless matter and we can now discuss the Seiberg duality in the modified geometry.
We have the situation of \cite{giku1} and the Seiberg duality then proceeds as an usual flop where 
\begin{equation}
C^3(\tilde{X},\tilde{Y}=Y,\tilde{Z})~~~\leftrightarrow~~~C^3(\tilde{Y}',\tilde{X}',\tilde{Z}'). 
\end{equation}

Now, if we want to see the effect of the flop in the original complex structure, we need to rotate back from the 
tilde coordinates to the original ones. This will leave the $\tilde{Y}'$ axis invariant but will change the $\tilde{X'}$ axis and 
$\tilde{Z'}$ axis. The compact North Pole of the ${\bf P^1}$ cycle is moved in its original position and its transition function is again   $Z' = 1/Z$.

But the rotation of the line bundle does not affect the non compact 2-cycle which remains in the rotated complex structure and it is 
not holomorphic in the original coordinates.  When the rotation starts, there is an
angle between the compact D5-branes and the non compact D5-branes and this tilting determines a tachyon to appear between the two stacks of branes. 
There are two ways to cancel the tachyon:

1) bind and rotate together the stacks of D5-branes

2) distance them such that the open string between them has no tachyonic mode. 

What is the tachyonic mass? It is related to the angle of rotation of the compact cycle with respect to the non compact cycle as
\begin{equation}
m^{2}_{\mbox{tach}}= - \frac{\mbox{tan}(\psi)}{l_s^2}
\end{equation}
where $\mbox{tan}(\psi)=\frac{\mu}{L_n}$ and $L_n$ is the distance to the cut-off beyond where the normal deformations of the 
cycle inside the Calabi-Yau is frozen \cite{agava2}.    

The final configuration is then obtained by combining the color D5-branes with some of the D5 branes on the non compact holomorphic 2-cycle 
ending on the $Y$ to give D5-branes on a  non compact holomorphic 2-cycle ending on the $X'$. 
The other cycle is the non compact non holomorphic 2-cycle ending on the $Y$ line of singularity.

After the duality and recombination of branes, the configuration with $M=0, q = \tilde{q} = 0$ is actually never obtained in the magnetic theory. This is because the
branes recombine before rotating back to the original geometry. In the magnetic theory we can turn on vevs for the 
field $M$. This means displacing the non compact cycle on the 
corresponding line bundle. If the non compact cycle is infinitesimally displaced on the line bundle, there is an attractive force which determines a bound between 
the branes. If the distance is bigger, than the stacks of D5 branes tend to reject each other and the theory goes to the 
SUSY vacua.

One can lift this configuration to F theory. 
We start with the explicit metric for the resolved conifold, obtained in \cite{bdt5}:
\begin{equation}
ds^2_{\cal M}~ = ~{\cal C}(r)^2~dr^2 +
{\cal C}(r)^{-2}~\Big(dz + Q~{\rm cot}~\theta_1~ dx + Q~{\rm
cot}~\theta_2~dy\Big)^2~ + {\cal
C}(r)~(d\theta_1^2 + dx^2) + {\cal C}(r)~(d\theta_2^2 + dy^2).
\end{equation}
where the compact cycle is in the $(y,\theta_2)$. The deformation of the geometry is obtained by considering a  rotation in the $(\theta_1,\theta_2)$ plane.
One can study the corresponding SUGRA solution in detail to derive the form of the non holomorphic cycles.

We can also lift to M theory where we encounter the problems in getting a solution for the single M5 brane pointed out in \cite{bena}. The above discussion 
hints that more general geometrical deformation from ${\cal N}=2$ theories to ${\cal N}=1$ theories 
are needed. We use arguments similar to the ones of \cite{dv1}. 
Start with an ${\cal N} = 2$ geometry $O(-2) + O(0)$ over a  ${\bf P^1}$ cycle and denote the direction $O(-2)$ and $O(0)$ by 
$\phi(z)$ and $\chi(z)$ respectively. The holomorphic Chern-Simons theories is described by the following action
\begin{equation}
\label{single}
S = \frac{1}{g_s} \int Tr(\phi \bar{D} \chi)
\end{equation}
If one turns on the Higgs fields $\phi(z)$ or $\chi(z)$, the ${\bf P^1}$ cycle is deformed into a non holomorphic curve 
$\cal{C}$ and the above action modifies into
\begin{equation}
S =  \frac{1}{g_s} \int_{Y} \Omega
\end{equation}
where $Y$ is a 3-chain containing  ${\bf P^1}$ and $\cal{C}$. 
By adding a superpotential for the field $\chi$ we deform the complex structure such that $\cal{C}$ becomes  ${\bf P^1}$ in the new complex structure. 
The complex structure of  the ${\cal N}=2$ theories is modified by perturbing the operator $\bar{\partial}_{\bar{j}}$ by
\begin{equation}
\bar{D} = \bar{\partial}_{\bar{j}} + A^{i}_{\bar{j}} \partial_{i}
\end{equation}
where $ A^{i}_{\bar{j}}$ is anti holomorphic one form and is related to the above superpotential \cite{kachru}.

The case of ADE quivers has been considered in \cite{dv2}. For the $A_n$ singularities, the equation (\ref{single}) is generalized to 
\begin{equation}
S = \frac{1}{g_s} \int Tr(\phi_i \bar{D} \chi_i),~i=1,\cdots,n
\end{equation}
There are $n$  ${\bf P^1_i}$ cycles which can be deformed into $n$ non holomorphic curves ${\cal{C}}_i$ by turning on the Higgs fields. 
By adding a superpotential for the fields $\chi_i$, we can reach the geometries:

a) all the  non holomorphic curves ${\cal{C}}_i$ change into  $n$  intersecting ${\bf P^1_i}$ cycles in the new geometry \cite{dv2}. 

b)  all the  non holomorphic curves ${\cal{C}}_i$ change into  $n$  ${\bf P^1_i}$ cycles, some of which do not intersect. This is the geometry corresponding to the 
electric theory.

c) some of the  non holomorphic curves ${\cal{C}}_i$ remain non holomorphic after adding the superpotential. This is the geometry 
corresponding to the magnetic theory.

The geometries of type b) and c) are taken into each other after Seiberg dualities. One can use the powerful duality with the matrix models in order to get 
more insights into the metastable vacua of \cite{iss}.

\section*{Acknowledgements}

We would like to thank Keshav Dasgupta, Radu Roiban and Angel Uranga for comments and Cumrun Vafa for helping us clarify in section 2.

\end{document}